\documentclass{webofc}
\usepackage[varg]{txfonts}   
\usepackage{color}
\usepackage{float}
\begin{document}
\title{\boldmath Long-distance QCD effects in FCNC $B\to \gamma l^+l^-$ decays}
%
%
\author{\firstname{Anastasiia} \lastname{Kozachuk}\inst{1,2}\fnsep\thanks{\email{anastasiia.kozachuk@cern.ch}} \and
        \firstname{Dmitri} \lastname{Melikhov}\inst{2,3,4}\fnsep\thanks{\email{dmitri_melikhov@gmx.de}} \and
        \firstname{Nikolai} \lastname{Nikitin}\inst{1,2,4}\fnsep\thanks{\email{nikolai.nikitine@cern.ch}}
}

\institute{M.~V.~Lomonosov Moscow State University, Faculty of Physics, 119991 Moscow, Russia
\and 
D.~V.~Skobeltsyn Institute of Nuclear Physics, M.~V.~Lomonosov Moscow State University, 119991 Moscow, Russia
\and
Institute for High Energy Physics, Austrian Academy of Sciences, Nikolsdorfergasse 18, A-1050 \\
Vienna, Austria 
\and 
Faculty of Physics, University of Vienna, Boltzmanngasse 5, A-1090 Vienna, Austria
\and
A.~I.~Alikhanov Institute for Theoretical and Experimental Physics, 117218 Moscow, Russia
}

\abstract{
This presentation reviews the main results of our recent work \cite{Kozachuk:2017mdk} on rare radiative leptonic 
decays $B_{d,s}\to\gamma\mu^+\mu^-$ and $B_{d,s}\to\gamma e^+e ^-$
induced by flavour-changing neutral currents (FCNC) in the Standard Model. 
}
\maketitle
%
Rare FCNC decays of $B$-mesons are forbidden at the tree level in the Standard Model and occur only via loop diagrams. Respectively, their 
branching ratios are very small, of order $10^{-8}-10^{-10}$ \cite{Ali:1996vf}. 
New particles can propagate in the loops and hence such processes are expected to be sensitive to the possible effects of New Physics. 
Several FCNC decays have been observed experimentally, exhibiting a few deviations from the Standard Model 
at the level of $2-3$ $\sigma$ (see discussion in \cite{Guadagnoli:2016erb,diego2017}). 

An appropriate framework for the theoretical description of FCNC $B$-decays is the effective field theory:
virtual heavy particles of the SM with the masses much greater than $m_b$ (i.e., $W$, $Z$, and $t$-quark) are integrated out 
yielding the $b\to q$ effective Hamiltonian \cite{Grinstein:1988me,Burasa,Burasb} 
\begin{eqnarray}
\label{Heff}
H_{\rm eff}^{b\to q}=\frac{G_F}{\sqrt{2}}V^*_{tq}V_{tb}\sum_i C_i(\mu) {\cal O}_i^{b\to q}(\mu).   
\end{eqnarray}
The basis operators ${\cal O}_i^{b\to q}(\mu)$ contain only dynamical light degrees of freedom ($u$, $d$, $s$, $c$, and $b$-quarks, leptons, 
photons and gluons). These light particles appear as dynamical degrees of freedom in the diagrams for the amplitudes of rare FCNC $B$-decays. 
The Wilson coefficients $C_i(\mu)$ absorb the contributions of the heavy particls ($W$, $Z$, and $t$ in the SM) given by the 
box and the penguin digrams; taking account of hard gluon exchanges in the Feynman diagrams leads to the dependence of $C_i$ on the scale $\mu$. 

Finally, the amplitude of the radiative leptonic $B$-decay is given by 
\begin{eqnarray} 
A(B_q\to \gamma ll)= \langle\gamma ll|H_{\rm eff}^{b\to q}|B_q\rangle.  
\end{eqnarray}
The presence of the $B$-meson in the initial state leads to the necessity of treating nonperturbative QCD effects; the amplitudes of 
FCNC rare leptonic $B$-decays involve a great variety of such nonperturbative QCD contributions. 


\noindent
\section*{Top-quark contributions}

\begin{figure}[ht]
\sidecaption
\centering
\includegraphics[width=5cm,clip]{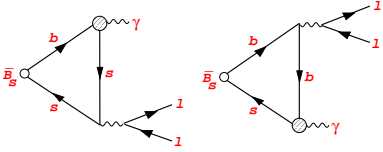}
\caption{Real photon emission from the penguin FCNC vertex.}
\label{fig-21}      
\end{figure}

\noindent
The diagrams generated by the $t$-quarks (as well as other heavy particles of the SM) in the loops are presented in Fig.~\ref{fig-2} and Fig.~\ref{fig-21}. 
The $B$-decay amplitudes corresponding to these diagrams are described in terms of $B\to \gamma$ transition form factors: 
\begin{figure}[ht]
\sidecaption
\centering
\includegraphics[width=9cm,clip]{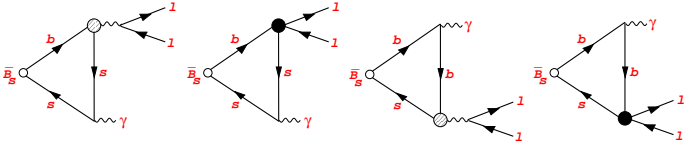}
\caption{Real photon emission from the valence quark.}
\label{fig-2}      
\end{figure}

\noindent
The diagrams of Fig.~\ref{fig-2} yield contributions that have no singularities in the physical decay region. 
For these contributions, we made use of the dispersion approach based on the relativistic constituent quark picture \cite{ma,mb,melikhov,msa,ms}: 
within this approach, the form factors are given by relativistic spectral representations via the meson relativistic 
wave functions. The transition form factors from the dispersion approach satisfy all rigorous constraints known from QCD for these quantities 
\cite{mb,Kruger:2002gf}. The model parameters (i.e., the constituent quark masses and the hadron wave functions) 
have been fixed by the requirement to reproduce the known values of meson weak decay constants \cite{Kozachuk:2017mdk}. 
The left diagram of Fig.~\ref{fig-21} leads to the contributions to the form factors that do have singularities (resonances and cuts) in the physical decay region. 
To calculate these contributions, we combined the results from the dispersion approach at $q^2<0$ with the gauge-invariant version  
of vector meson dominance \cite{Nachtmann}. 

\begin{figure}[ht]
\sidecaption
\centering
\includegraphics[width=6.5cm,clip]{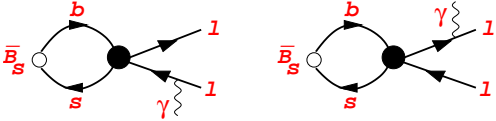}
\caption{The Bremsstrahlung contribution.}
\label{fig-22}      
\end{figure}

Finally, Fig.~\ref{fig-22} displays the Bremsstrahlung contribution to the $B_{s,d}\to\gamma\ell^+\ell^-$ amplitude; 
it is given in terms of the $B$-meson decay constant $f_B$ and is proportional to the mass of the lepton in the final state \cite{Melikhov:2004mk}.  

\section*{Charm-quark contributions}

\begin{figure}[ht]
\sidecaption
\centering
\includegraphics[width=3.5cm,clip]{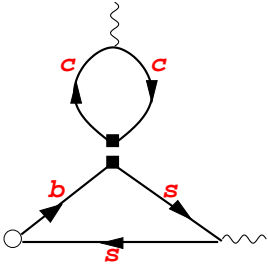} \hspace{1cm}\includegraphics[width=3.5cm,clip]{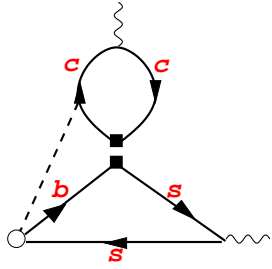}
\caption{Factorizable (left) and nonfactorizable (right) charming loops. Dashed line in the right figure stands for the soft gluon.}
\label{fig-5}      
\end{figure}

\noindent
Whereas heavy degrees of freedom ($t$, $W$, $Z$) are described by the effective Hamiltonian, 
light degrees of freedom, in particular $c$ and $u$ quarks, remain dynamical and their contributions in the loops 
should be taken into account.
\begin{figure}[ht]
\sidecaption
\centering
\includegraphics[width=8cm,clip]{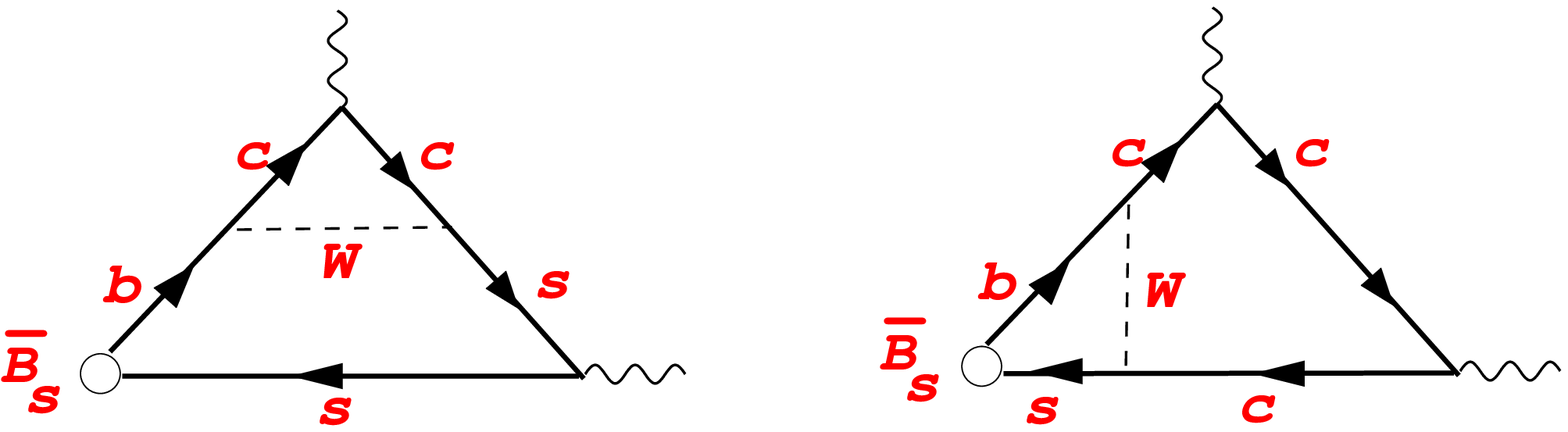}
\caption{Diagrams with c-quarks in the loops: penguin diagram (left), weak annihilation diagram (right)}
\label{fig-3}      
\end{figure}
The diagrams for the charm-loop contributions are presented in Fig.~\ref{fig-3}: 
the numerically dominant penguin diagram (left) and the weak-annihilation diagram (right). When the $W$-boson line is shrinked to a point, and the soft-gluon exchanges 
between the charm-loop and the $B$-meson loop are taken into account, the charming penguin is reduced to two classes of diagrams shown in Fig.~\ref{fig-5}.  

The structure of the charm-loop contributions to the $B_s\to \gamma l^+l^-$ amplitude has the form:
\begin{eqnarray}
\label{H1}
&&H_{\mu\alpha}(k',k)=
-\frac{G_F}{\sqrt{2}}V_{cb}V^*_{cs}e\Bigg[
\epsilon_{\mu\alpha k'k}H_V
-i\left(g_{\alpha\mu}\,kk'-  k'_\alpha k_\mu\right)H_A\Bigg],  
\end{eqnarray} 
with $0 <q^2< M_B^2$, including the region of the charmonium resonances. Perturbative QCD cannot be applied here and non-perturbative 
approaches based on hadron degrees of freedom are necessary, see discussion in \cite{ali,bbns_duality,hidr,cc_1a,cc_1b,cc_1c,cc_1d,cc_2}. 
For $H_i(q^2,0)$ one may write dispersion representation in $q^2$ with two subtractions \cite{hidr}:
\begin{eqnarray}
\label{Hdisp}
H_i(q^2,0)=a_i+ b_i q^2+(q^2)^2\left\{ \sum_{\psi=J/\psi,\psi'} \frac{f_\psi {\cal A}^{i}_{B\psi\gamma}}{m_\psi^3(m_\psi^2-q^2-i m_\psi \Gamma_\psi)}+
h_i(q^2)\right\},\quad i=V,A,   
\end{eqnarray} 
where the functions $h_i(q^2)$ describe the hadron continuum including the 
broad charmonium states lying above the $DD$ threshold. At $q^2>4M_D^2$, we take into account the known broad 
vector $\psi_n$ ($n=3,\dots,6$) resonances. The subtraction constants $a_i$ and $b_i$ are determined by matching to the known results 
from light-cone sum rules at $q^2\le 4m_c^2$, including non-factorizable corrections calculated in \cite{hidr}.
The absolute values of the amplitudes ${\cal A}^{i}_{B\psi\gamma}$ may be measured in $B\to \psi_i\gamma$ decays. 
It should be understood, however, that nonfactorizable gluons may generate
complex relative phases between the contributions of different charmonia \cite{cc_1b}.  
These possible non-universal (i.e., process-dependent and thus in general different in $B\to \gamma ll$ and $B\to V ll$) 
relative phases cannot be determined by pQCD-based calculations and provide one of 
the main sources of the theoretical uncertainties for rare radiative leptonic decays. 
Further details and subtleties see \cite{Kozachuk:2017mdk,melikhovkozachuk}. 

\section*{Results}

\begin{figure}[ht]
\begin{center}
\includegraphics[width=6cm]{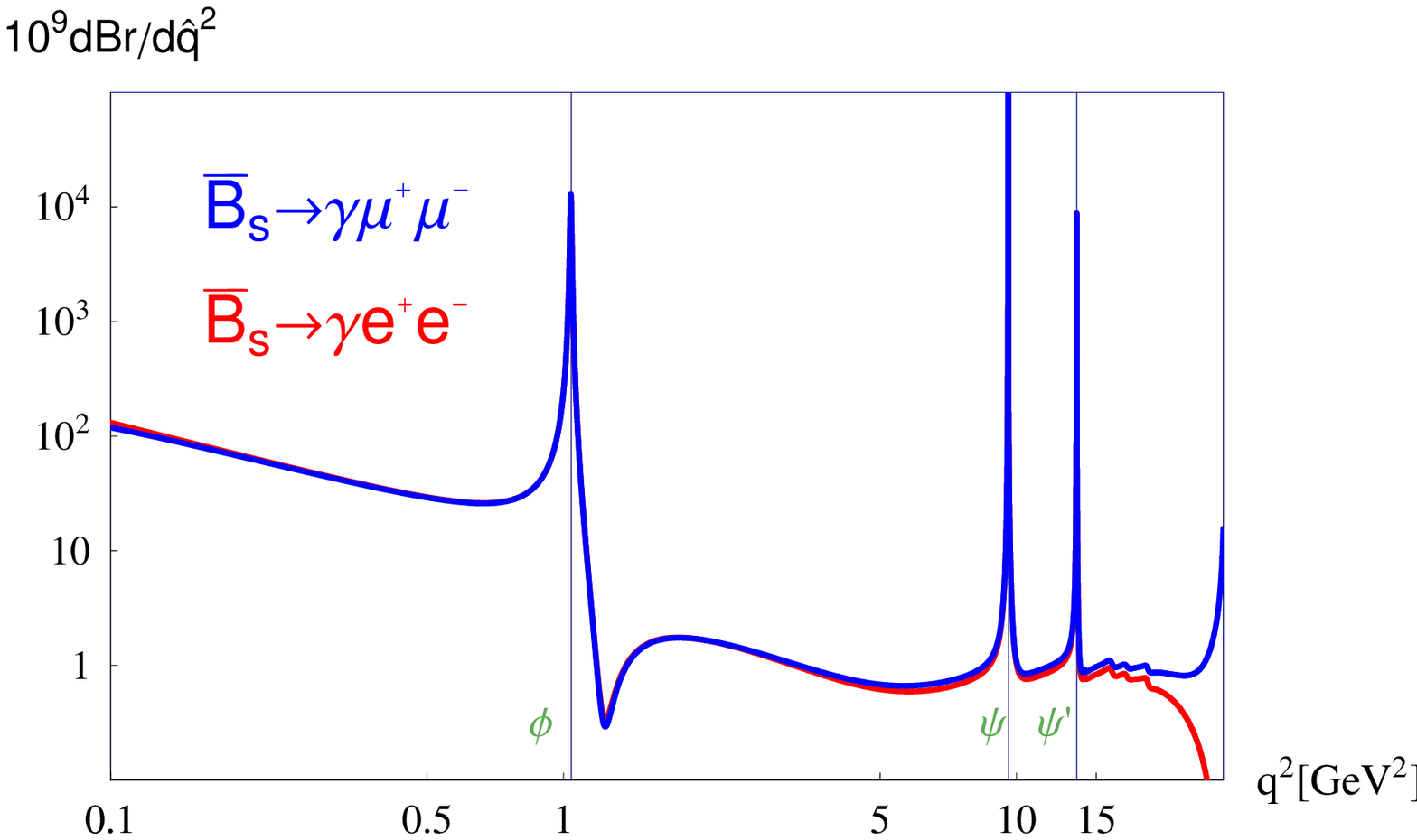}
\vspace{.4cm}
\includegraphics[width=6cm]{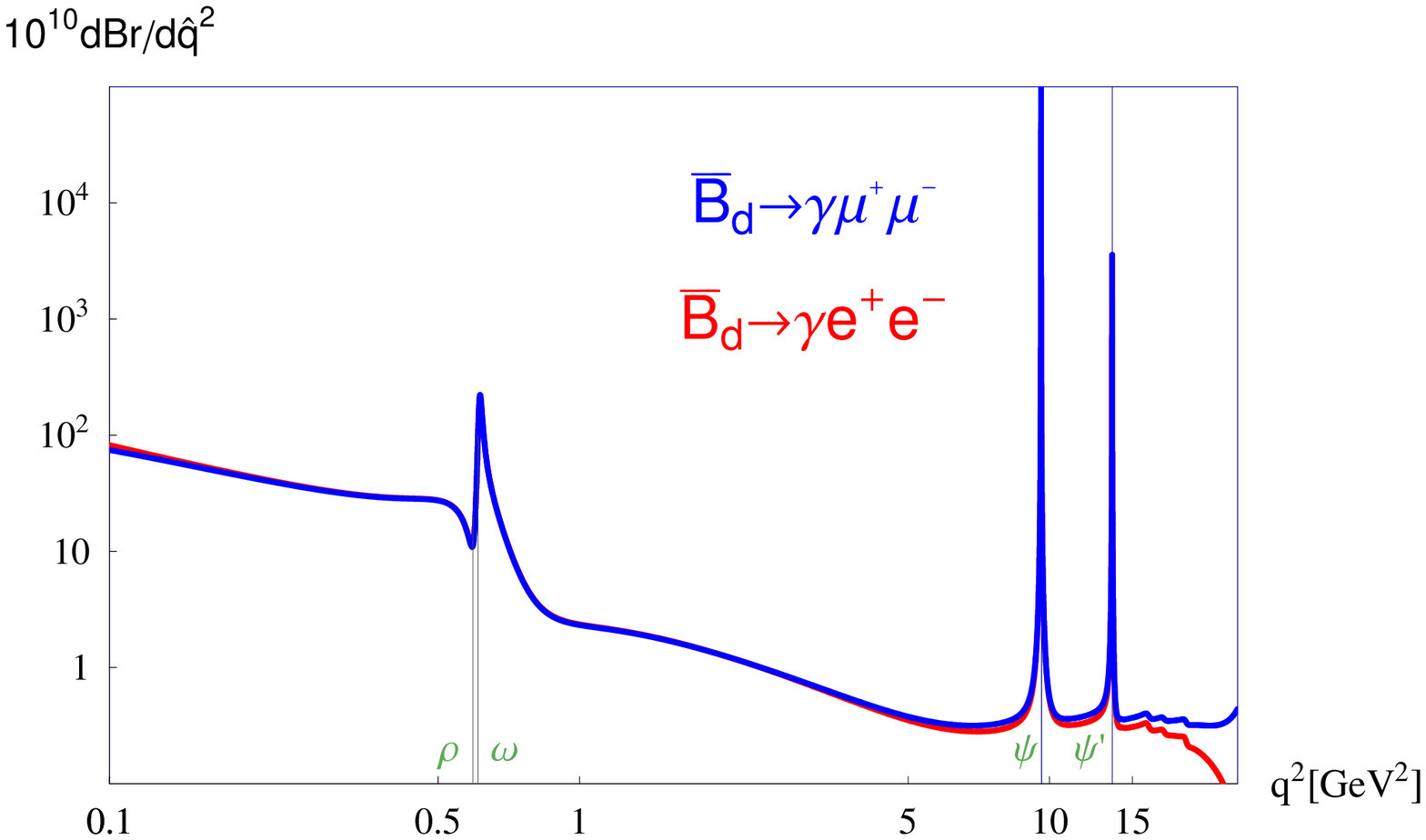}
\caption{Differential branching fractions for $B_s\to \gamma l^+l^-$ (left) and $B_d \to \gamma l^+l^-$ (right) decays. 
Blue lines - $\mu^+\mu^-$ final state, red lines - $e^+e^-$ final state.}
\label{Plot:BR} 
\end{center}
\end{figure}

\noindent
The differential branching ratios are shown in Fig.~\ref{Plot:BR}. 
The plots in Fig.~\ref{Plot:BR} correspond to the description of the charm-loop effects according to Eq.~(\ref{Hdisp}), and further assuming that 
all charmonia contribute with the same positive sign (coinciding with the sign of the factorizable contribution). 

In the region $q^2\le 6$ GeV$^2$, the charming loops provide a small contribution at the level of a few percent, so  
the distributions in this region may be predicted with a good accuracy, mainly limited by the form-factor uncertainty. 
Our estimates \cite{Kozachuk:2017mdk} read
\begin{eqnarray}
\label{Br1-6GeV2}
&&{\cal B}(\bar B_s\to \gamma l^+l^-)|_{q^2\in[1,6]\, {\rm GeV}^2}=(6.01\pm 0.08)10^{-9}\nonumber\\
&&{\cal B}(\bar B_d\to \gamma l^+l^-)|_{q^2\in[1,6]\, {\rm GeV}^2}=(1.02\pm 0.15)10^{-11}.
\end{eqnarray}

\begin{figure}[ht]
\centering
\includegraphics[width=6cm,clip]{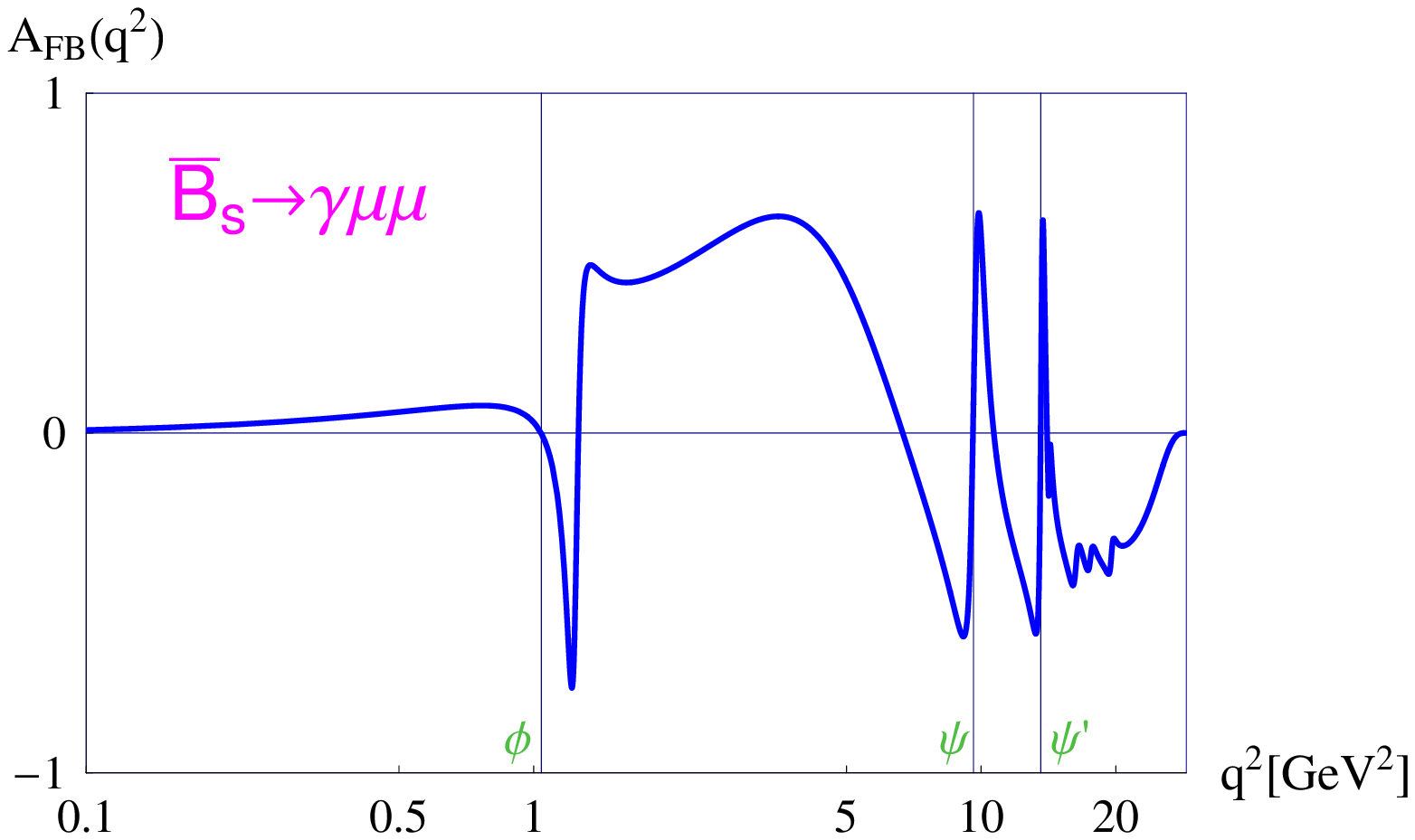}
\hspace{.4cm}
\includegraphics[width=6cm,clip]{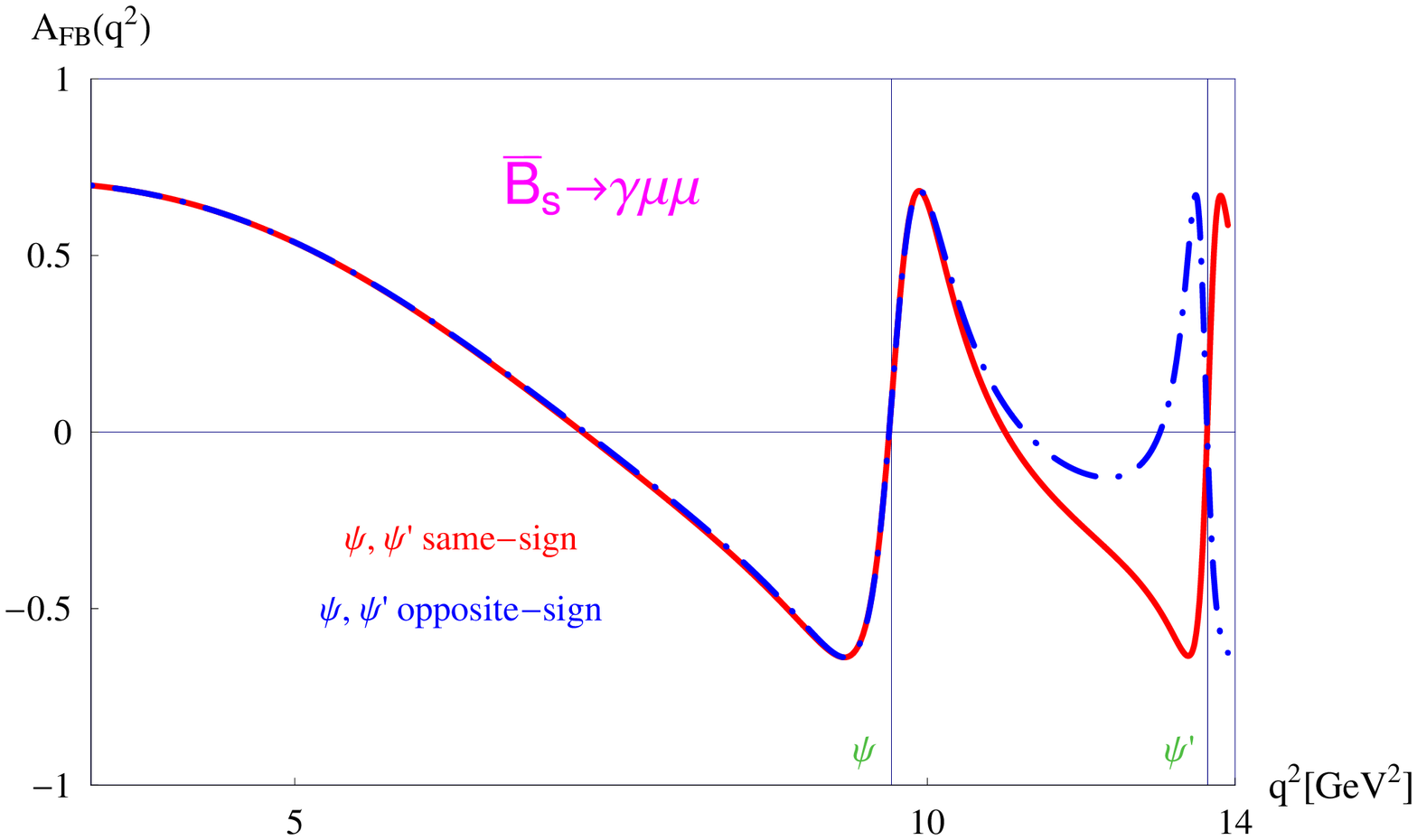}
\caption{Forward-backward asymmetry for $B_s\to \gamma\mu\mu$ decays. The left plot shows the asymmetry for all $q^2$ for the contributions of 
all charmonia taken of the same positive sign, equal to that of the factorizable contribution; the right plot shows the sensitivity 
of $A_{\rm FB}$ to the relative signs of $\psi$ and $\psi'$: 
solid (red) line - both signs positive, dashed (blue) line - the $\psi$ contribution taken positive, 
whereas the $\psi'$ contribution negative.}
\label{fig-6}       
\end{figure}

\noindent
For the $B_s\to \gamma l^+l^-$ transition, the dominant contribution is given by the $\phi$-meson. 
Its parameters are known well, leading to the small uncertainty in the $B_s\to \gamma l^+l^-$ 
decay rate integrated over the range of $q^2=[1,6]$ GeV$^2$. For the $B_d\to \gamma l^+l^-$ transition, 
the known contribution of the vector resonances is less important, and the branching ratio uncertainty 
reflects to a large extent the 10\% uncertainty in the form factor contributions given by Fig.~\ref{fig-2}.

The $A_{FB}$ for $\bar B_s\to \gamma\mu\mu$ is shown in Fig.~\ref{fig-6}. The right plot exhibits the sensitivity of $A_{FB}$ 
in the region between $\psi$ and $\psi'$ to the relative signs of the contributions of these states.

Further results and details see our paper \cite{Kozachuk:2017mdk}. 


\section*{Acknowledgments}
\vspace{-.1cm}

\noindent
D.~M. thanks the Organizers of QCD@work for their warm hispitality in Matera and FWF for finacial support under project P29028. 
The results given in Eq.~(4) were obtained with support of grant 16-12-10280 of the Russian Science Foundation (A.~K. and N.~N.).



\begin{thebibliography}{}

\vspace{-0.1cm}
\bibitem{Kozachuk:2017mdk} 
A.~Kozachuk, D.~Melikhov and N.~Nikitin,
{\it Rare FCNC radiative leptonic $B_{s,d}\to \gamma l^+l^-$ decays in the standard model},
Phys.\ Rev.\ D {\bf 97}, no. 5, 053007 (2018) [arXiv:1712.07926]
\bibitem{Ali:1996vf}
A.~Ali,
{\it $B$ decays, flavor mixings and CP violation in the SM},
[hep-ph/9606324]. 
%
\bibitem{Guadagnoli:2016erb}
D.~Guadagnoli, D.~Melikhov and M.~Reboud,
{\it More Lepton Flavor Violating Observables for LHCb's Run 2}, 
Phys.\ Lett.\ B {\bf 760}, 442 (2016) [arXiv:1605.05718].
%
\bibitem{diego2017}
D.~Guadagnoli,
{\it Flavor anomalies on the eve of the Run-2 verdict}, 
Mod.~Phys.~Lett. A{\bf 32}, 1730006 (2017) [arXiv:1703.02804]. 
%
\bibitem{Grinstein:1988me}
B.~Grinstein, M.~J.~Savage and M.~B.~Wise,
{\it  $B\to X(s) e^+e^-$ in the Six Quark Model},
Nucl.\ Phys.\ B {\bf 319} 271 (1989). 

\bibitem{Burasa} 
A.~J.~Buras and M.~Munz,
{\it Effective Hamiltonian for $B\to X(s) e^+e^-$ beyond leading logarithms in the NDR and HV schemes},
Phys.\ Rev.\ D {\bf 52}, 186 (1995) [hep-ph/9501281]. 

\bibitem{Burasb} 
G.~Buchalla, A.~J.~Buras, and M.~E.~Lautenbacher, 
{\it Weak decays beyond leading logarithms}, Rev.~Mod.~Phys.~ {\bf 68}, 1125 (1996) [hep-ph/9512380].

\bibitem{ma} 
D.~Melikhov, 
{\it Form-factors of meson decays in the relativistic constituent quark model},  
Phys.\ Rev.\ D {\bf 53}, 2460 (1996) [hep-ph/9509268]. 

\bibitem{mb}
D.~Melikhov, 
{\it Heavy quark expansion and universal form-factors in quark model}, 
Phys.\ Rev.\ D {\bf 56}, 7089 (1997) [hep-ph/9706417]. 

\bibitem{melikhov} 
D.~Melikhov,  
{\it Dispersion approach to quark binding effects in weak decays of heavy mesons}, 
Eur.~Phys.~J.~direct {\bf C4}, 2 (2002) [hep-ph/0110087]. 

\bibitem{msa} 
M. Beyer and D. Melikhov, {\it Form-factors of exclusive $b\to u$ transitions}, 
Phys. Lett. B {\bf 436}, 344 (1999) [hep-ph/9807223]. 

\bibitem{ms} 
D. Melikhov and B. Stech, 
{\it Weak form-factors for heavy meson decays: An Update}, 
Phys. Rev. D {\bf 62}, 014006 (2000) [hep-ph/0001113]. 

\bibitem{Kruger:2002gf} 
F.~Kruger and D.~Melikhov,
{\it Gauge invariance and form-factors for the decay $B\to \gamma l^+ l^-$}, 
Phys.\ Rev.\ D {\bf 67}, 034002 (2003) [hep-ph/0208256].
%
\bibitem{Nachtmann} 	
D.~Melikhov, O.~Nachtmann, V.~Nikonov and T.~Paulus, 
{\it Masses and couplings of vector mesons from the pion electromagnetic, weak, and pi gamma transition form-factors}, 
Eur.~Phys.~J.~{\bf C34}, 345 (2004) [hep-ph/0311213].

\bibitem{Melikhov:2004mk} 
D.~Melikhov and N.~Nikitin,
{\it Rare radiative leptonic decays $B(d,s)\to \gamma l^+ l^-$}, 
Phys.\ Rev.\ D {\bf 70}, 114028 (2004) [hep-ph/0410146]. 
%
\bibitem{ali}
A.~Ali, T.~Mannel, T.~Morozumi, 
{\it Forward backward asymmetry of dilepton angular distribution in the decay $b\to s l^+l^-$}, 
Phys.\ Lett. \ B {\bf 273}, 505 (1991). 

\bibitem{bbns_duality}
M.~Beneke, G.~Buchalla, M.~Neubert, and C.~T.~Sachrajda, 
{\it Penguins with Charm and Quark-Hadron Duality}, 
Eur.~Phys.~J. C{\bf 61}, 439 (2009) [arXiv:0902.4446].

\bibitem{hidr}
A.~Khodjamirian, T.~Mannel, A.~Pivovarov, and Y.-M.~Wang, 
{\it Charm-loop effect in $B\to K^{(*)} l^+l^-$ and $B\to K^*\gamma$}, 
JHEP {\bf 09}, 089 (2010) [arXiv:1006.4945]. 

\bibitem{cc_1b}
J.~Lyon and R.~Zwicky, 
{\it Resonances gone topsy turvy - the charm of QCD or new physics in $b\to sl^+l^-$?},  
[arXiv:1406.0566]. 

\bibitem{cc_1a}
M.~Ciuchini, M.~Fedele, E.~Franco, S.~Mishima, A.~Paul, L.~Silvestrini, and M.~Valli, 
{\it $B\to K^*l^+l^-$ decays at large recoil in the Standard Model: a theoretical reappraisal}, 
JHEP {\bf 1606}, 116 (2016) [arXiv:1512.07157]. 

\bibitem{cc_1c}
S.~Brass, G.~Hiller, and I.~Nisandzic, 
{\it Zooming in on  $B\to K^{(*)} l^+l^-$ decays at low recoil }, 
Eur.~Phys.~J. C{\bf 77}, 16 (2017) [arXiv:1606.00775]. 

\bibitem{cc_1d}
LHCb Collaboration, R.~Aaij {\it et al.},
{\it Measurement of the phase difference between short- and long-distance amplitudes in the $B^+\to K^{+} \mu^+\mu^-$ decay}, 
Eur.~Phys.~J. C{\bf 77}, 161 (2017) [arXiv:1612.06764].  

\bibitem{cc_2}
T.~Blake, U.~Egede, P.~Owen, G.~Pomery, K.~Petridis,
{\it An empirical model of the long-distance contributions to $\bar B^0\to \bar K^{*0} \mu^+\mu^-$}
[arXiv:1709.03921].
%
\bibitem{melikhovkozachuk} 
A.~Kozachuk and D.~Melikhov,
{\it Nonfactorizable charm-loop effects in rare FCNC $B$-decays},
[arXiv:1805.05720].
%

\end{thebibliography}
\end{document}